\documentclass{article}
\usepackage{spconf,amsmath,graphicx}

\usepackage{graphicx}
\usepackage{color}
\usepackage{amssymb,amsmath,bm}
\usepackage{textcomp}
\usepackage{tikz}
\usepackage[utf8]{inputenc}
\usepackage{pgfplots}
\usepackage{url}

\usepackage{bm}

\title{Speaking style adaptation in Text-To-Speech synthesis using Sequence-to-sequence models with attention}
\name{Bajibabu Bollepalli \qquad Lauri Juvela \qquad Paavo Alku}
\address{Department of Signal Processing and Acoustics, Aalto University, Finland}

\begin{document}
\sloppy
\ninept
\maketitle
\begin{abstract}
   Currently, there are increasing interests in text-to-speech (TTS) synthesis to use sequence-to-sequence models with attention. These models are end-to-end meaning that they learn both co-articulation and duration properties directly from text and speech. Since these models are entirely data-driven, they need large amounts of data to generate synthetic speech with good quality. However, in challenging speaking styles, such as Lombard speech, it is difficult to record sufficiently large speech corpora. Therefore, in this study we propose a transfer learning method to adapt a sequence-to-sequence based TTS system of normal speaking style to Lombard style. Moreover, we experiment with a WaveNet vocoder in synthesis of Lombard speech.  
   We conducted subjective evaluations to assess the performance of the adapted TTS systems. The subjective evaluation results indicated that an adaptation system with the WaveNet vocoder clearly outperformed the conventional deep neural network based TTS system in synthesis of Lombard speech. 
  
 \end{abstract}
\begin{keywords}
 TTS, Tacotron, Lombard style, Adaptation
\end{keywords}
\section{Introduction} \label{sec:intro}
    For better communication, humans typically change their speaking style depending upon the acoustic and auditory environment. In noisy surroundings, humans adapt to Lombard style \cite{lombard1911} in order to improve speech intelligibility. When compared to speech of normal style, Lombard speech differs in many respects both acoustically and phonetically by exhibiting, for example, an increase in vocal intensity and fundamental frequency ($F0$), a decrease in spectral tilt, and an increased prominence in the production of vowels compared to consonants \cite{junqua1993lombard,lu2008speech}.
    
    In literature, it has been shown that intelligibility of synthetic normal speech is significantly lower than that of synthetic Lombard speech when evaluated in noisy surroundings \cite{cooke2013evaluating}. Use of text-to-speech (TTS) systems, however, is becoming increasingly prevalent in real-life noisy environments such as in GPS navigation in busy traffic or public announcement systems in train and bus stations. Thus there is a great need for TTS systems to improve their speech intelligibility in noisy conditions by adapting the speaking style of the synthesis to Lombard. In this article, we investigate a transfer learning technique to adapt a sequence-to-sequence based normal speaking style TTS system to Lombard style\footnote{Samples available at \url{http://tts.org.aalto.fi/lombard_seq2seq/}}.

\section{Related work} \label{sec:related_work}
     In TTS, there are many studies on the adaptation of the speaking style, including Lombard, e.g., \cite{yamagishi2009analysis,raitio2014synthesis}. These previous studies are almost exclusively based on statistical parametric speech synthesis (SPSS) due to the technology's benefits in adaptation abilities and flexibility in changing voice characteristics. SPSS systems typically require a moderate amount of high-quality audio data per speaker to generate synthetic speech with good quality (e.g. 5 hours in \cite{qian2014training}). However, collecting several hours of speech data from one speaker is difficult, if not impossible, for high vocal effort speaking styles such as shouting and Lombard. To circumvent the data scarcity issue, adaptation techniques are usually employed in SPSS by allowing the use of a smaller amount of data for the particular speaking style to be synthesized. Speaking style adaptation of TTS was first used in hidden Markov model (HMM)-based SPSS systems. In adapting HMM-based systems, initial HMMs, trained on normal speech, were adapted by a small amount of Lombard speech using, for example, a technique called constrained structural maximum \emph{a posteriori} linear regression (CSMAPLR) \cite{yamagishi2009analysis}. In the more recent deep neural network (DNN) based SPSS systems, the adaptation can be done at three levels: 1) input level, 2) model level, and 3) output level \cite{wu2015study,Luong2018,huang2018linear,wu2018feature}. Previous studies have demonstrated that the naturalness of synthetic speech generated with DNN-based SPSS systems is higher than that of HMM-based systems \cite{ze2013statistical,fan2014tts}, and this applies also to adaptation to Lombard speech \cite{bollepalli2017lombard}.
 
    Even though promising results have been obtained in adaptation of synthetic speech using the SPSS framework, this conventional TTS paradigm has limitations that affect the synthesis naturalness. Conventional SPSS systems consists of two separate blocks, 1) front-end and 2) back-end. The front-end processes the input text by producing the numerical, linguistic representation \cite{zen2009statistical}. The back-end first maps the linguistic representation to acoustic features and then passes the mapped acoustic features to a vocoder to render the speech waveform. In this pipeline, both the front-end and back-end are usually constructed independently \cite{zen2009statistical}. Moreover, errors caused in each block can accumulate and degrade the overall performance of the system. Further, each block needs its own expertise to tune the system. 
    
    Recently, a more simplified framework using sequence-to-sequence models with attention was proposed for TTS \cite{sotelo2017char2wav,wang2017tacotron,arik2017deep}. These models depend heavily on encoder-decoder neural network structures that map a sequence of characters to a sequence of acoustic frames. These models combine the front-end and back-end and learn relations between them from data only. When sequence-to-sequence models are coupled with neural vocoders, they enable generating raw waveforms directly from text \cite{ping2018clarinet}. In \cite{tacotron2}, it was demonstrated that state-of-the-art results in TTS can be achieved with the sequence-to-sequence technology. Despite their success in producing high-quality synthetic speech, sequence-to-sequence systems, however, need a sizable amount of data (i.e. text/audio pairs). In  \cite{chung2018semi}, for example, it was concluded that around 10 hours of text/speech pairs are needed to get decent quality in synthetic speech by a sequence-to-sequence model such as Tacotron \cite{wang2017tacotron}. From now on, we denote sequence-to-sequence model based TTS as Seq2Seq-TTS.
 
    In Seq2Seq-TTS, only a few investigations have studied methods to synthesize speech of different speakers using a limited amount of data \cite{ping2018deepvoice3,arik2018neural}. These studies have employed speaker embeddings, which contain speaker-specific characteristics for multispeaker speech synthesis. However, extracting the speaker embeddings for unseen speakers in training data may require a huge amount of data to train a separate speaker-encoder network \cite{jia2018transfer}. However, to learn \textit{style-specific} embeddings for challenging speaking styles does not call for having that much data. Thus, in this study we propose a method to effectively leverage an existing large volume of normal speech data to synthesize Lombard speech using a Seq2Seq-TTS system. Our study is to some extent similar to  \cite{chung2018semi} that recently proposed a semi-supervised technique to reduce the data requirements by utilizing freely available data.
    
    The contributions of the paper are twofold. First, we develop a speaking style adaptation system using a Seq2Seq-TTS model. Second, we study the use of a WaveNet vocoder \cite{Tamamori2017} for the application of Lombard speech synthesis. To the best of our knowledge, the current study is the first investigation on speaking style adaptation of speech synthesis using a modern Seq2Seq-TTS system.

\section{Seq2Seq-TTS system} \label{sec:seq2seq_tts}
   
\begin{figure}[htpb]
        \centering
        \includegraphics[width=0.9\linewidth, height=3.4cm]{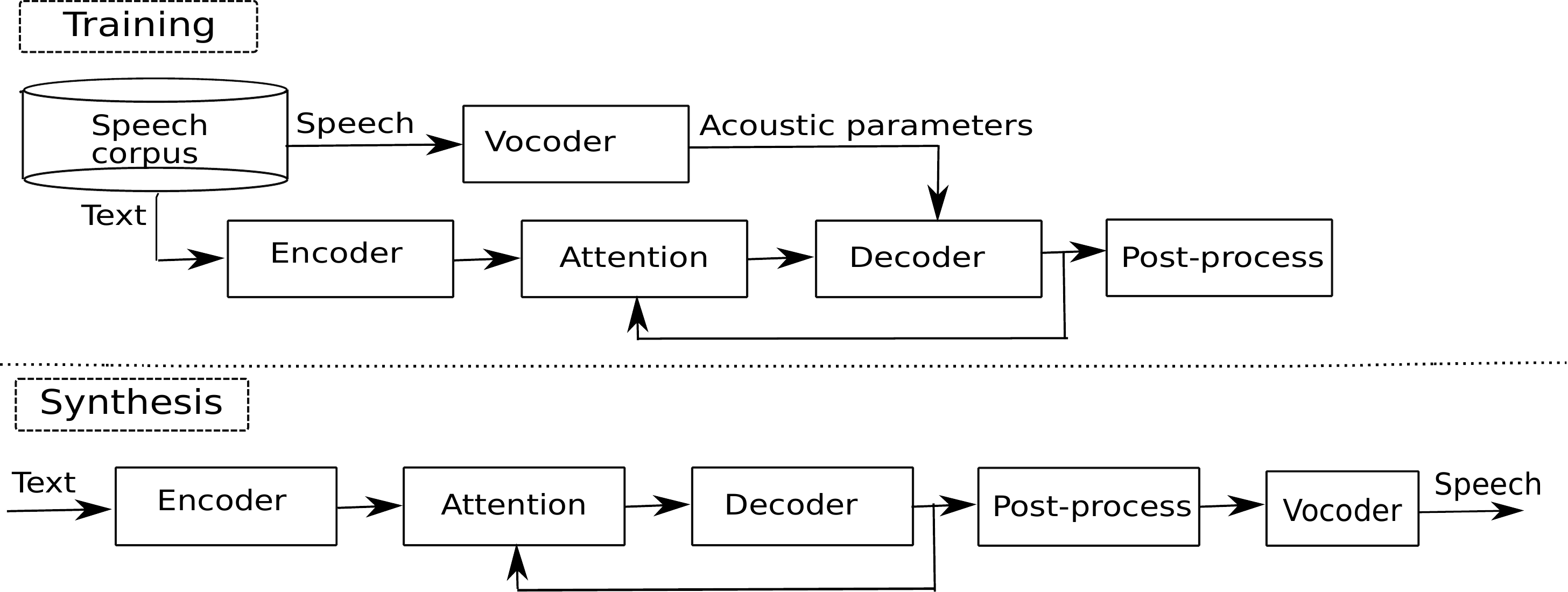}
        \caption{{\it General block diagram of a sequence-to-sequence based TTS system.}}
        \label{fig:seq2seq_block}
    \end{figure}

   
   Figure~\ref{fig:seq2seq_block} depicts a general block diagram of a TTS system using the sequence-to-sequence model with attention. The model accepts either mono-phonemes or graphemes as inputs and emits acoustic parameters as outputs. It consists of three main components: 1) encoder, 2) attention, and 3) decoder. The encoder takes text sequence $\bm{x}$ of length $L$ as input, which represented either in the character or phoneme domain as one-hot vectors. The encoder learns a continuous sequential representation $\bm{h}$ using various neural network architectures such as LSTMs \cite{wang2017tacotron,tacotron2} and/or CNNs \cite{ping2018deepvoice3}.
   \begin{equation}
       \bm h = \mathrm{encoder}(\bm x)
   \end{equation}
   At each output time step $t$, both the attention and decoder modules work together in the following manner:
   \begin{align}
       \alpha_t & = \mathrm{attention}( s_{t-1},  \alpha_{t-1}, \bm{h}) \\
        c_t & = \sum_{j=1}^L \alpha_{t,j} h_j \\
       y_t & = \mathrm{decoder}(s_{t-1}, c_t)
   \end{align}
    where $s_{t-1}$ is the $(t-1)$-th state of the decoder recurrent neural network and 
    $\alpha_t \in \mathbb{R}^L$ 
    are the attention weights or the alignment and $c_t$ is the context or attention vector. The decoder takes the previous hidden state $s_{t-1}$ and the current context vector $c_t$ as inputs and generates the current output $y_t$. This process runs until the end of the utterance is reached
    
    In order to synthesize the speech waveform, Seq2Seq-TTS systems use different vocoding approaches. Initial studies predict mel-spectrograms as output, mapping them to linear spectrograms and further to speech waveforms using the Griffin-Lim algorithm \cite{wang2017tacotron}. Recent studies, however, generate speech waveforms with the neural WaveNet vocoder, which is conditioned using the predicted mel-spectrograms \cite{tacotron2}. In this study, we predict the World vocoder \cite{morise2016world} parameters as well as mel-spectrograms as the system outputs, which are later used in conditioning the WaveNet vocoder to generate the final speech waveform.
    
    \subsection{Adaptation of Seq2Seq models} \label{subsec:adapt_seq2seq}
    Previous studies \cite{ping2018deepvoice3,nachmani2018fitting} have showed that Seq2Seq models can be adapted to different speakers by adding the speaker embeddings as an input to the model. In addition, some TTS studies \cite{jia2018transfer,arik2018neural} have used the fine-tuning technique to adapt the system to new speakers. In this approach, all the parameters of the network are fine-tuned by the new speaker data. In \cite{chung2018semi}, which is the study closest to the current investigation, each component of the model is trained separately with freely available data. 
    Once the encoder and decoder are pre-trained with existing data, a fine-tuning method is used to train the whole model on a small data set. Results of \cite{chung2018semi} show that the model can generate synthetic speech of good quality using around 30 minutes of data. In the present study, we first train a Seq2Seq-TTS system using a large amount of normal speech of one speaker and then fine-tune the learned model with normal speech of another speaker with limited data. Finally, using Lombard speech of the latter speaker, we fine-tune the model again to synthesize Lombard speech. We predict both mel-spectrograms and the World vocoder parameters as output acoustic frames. To render final speech waveform, we employ both the WaveNet vocoder and the World vocoder.

    \begin{figure*}[ht]
    \begin{minipage}[b]{.24\linewidth}
      \centering
      \centerline{\includegraphics[width=4cm, height=5cm]{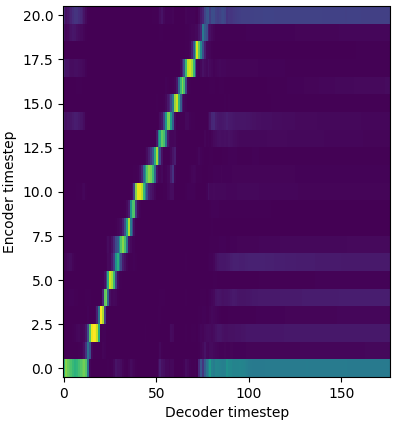}}
      \centerline{(a) Nick (normal) standalone}\medskip
    \end{minipage}
    \begin{minipage}[b]{.24\linewidth}
      \centering
      \centerline{\includegraphics[width=4cm, height=5cm]{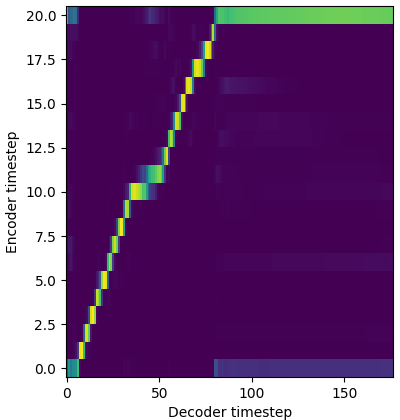}}
      \centerline{(b) Nancy (normal) standalone}\medskip
    \end{minipage}
    \begin{minipage}[b]{.24\linewidth}
      \centering
      \centerline{\includegraphics[width=4cm,height=5.0cm]{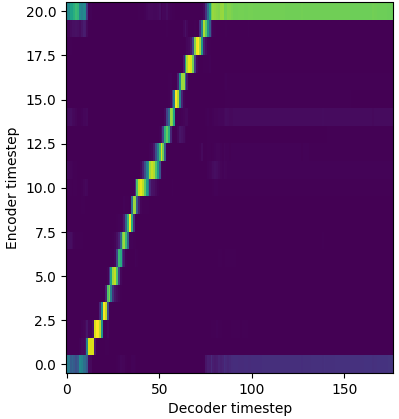}}
      \centerline{(c) Nick (normal) adapted}\medskip
    \end{minipage}
    \begin{minipage}[b]{.24\linewidth}
      \centering
      \centerline{\includegraphics[width=4cm,height=5.0cm]{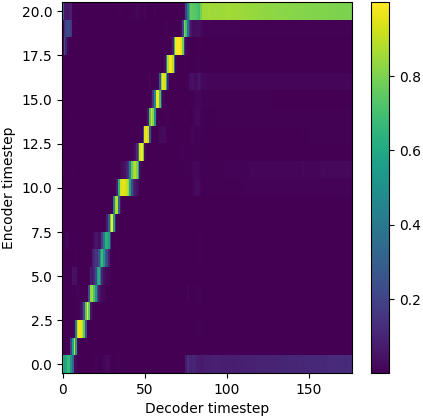}}
      \centerline{(d) Nick (Lombard) adapted}\medskip
    \end{minipage}
    \vspace{-3mm}
    \caption{Alignments obtained by the different systems, the sentence is \textsc{Paper will dry out when wet.} The x-axis and y-axis of each plot correspond to the mel-spectrograms of speech signal and the phonemes of the text, respectively.}
    \label{fig:align_plots}
    \end{figure*}

\section{Experiments} \label{sec:experiments}
    \subsection{Speech material} \label{subsec:speechmaterial}
    Our initial Seq2Seq-TTS model was trained on the Blizzard Challenge 2011 speech corpus \cite{blizzard2011}. The corpus contains around 12000 utterances (which adds up to around 16 hours) read by a US professional female voice talent named Nancy. We employed the Hurricane Challenge speech data \cite{hurricane} for adaptation to Lombard style. The Hurricane Challenge data was spoken by a British male voice professional named Nick. The Nick data consists of both normal and Lombard styles. The normal speech data consists of 2592 utterances (which adds up to 2 hours), and the Lombard speech data consists of 720 utterances (which adds up to 30 minutes). All the speech data was sampled with 16 kHz. The data was partitioned into train, valid and test sets as shown in Table \ref{tab:data_split}. 
 
    \begin{table}[htbp]
          \caption{Partition of the data (number of utterances) used in the present study.}
         \label{tab:data_split}
         \centering
            \begin{tabular}{|c|c|c|c|}
                \hline
                 Speaker (Style) & Train & Valid & Test  \\ \hline
                 Nancy (normal) & 11000   & 200 & 800  \\ \hline
                 Nick (normal) & 2400    & 72 & 120 \\ \hline
                 Nick (Lombard) & 500     &  100 & 120  \\ \hline
            \end{tabular}
     \end{table}

    \subsection{Systems built for comparison} \label{subsec:systems}
    Two types of acoustic parameters were extracted from the speech signals: 1) World vocoder parameters, and 2) mel-spectrograms. The World vocoder parameters consisted of mel-generalized cepstrum (MGC), fundamental frequency ($F0$) and band aperiodicity (BAP) with the dimensions 60, 1, and 1, respectively. The mel-spectrograms were extracted using the LibROSA \footnote{\url{https://github.com/librosa/librosa}} package and its dimension was 80. The World vocoder parameters were extracted at a 5-ms frame rate whereas the mel-spectrogram features were extracted at a 12.5-ms frame rate. The $F0$ values were linearly interpolated in unvoiced regions and transformed into the $log$-domain.
    
    We built a total of five systems for comparison as described in Table~\ref{tab:systems}. The systems were different in terms of their output parameter types and the vocoder used. System S1 is the baseline system which uses a LSTM-type of recurrent neural network (RNN)- based TTS system for adaptation, and synthesizes the speech waveform using the World vocoder. System S2 is built using the Seq2Seq-TTS model, and the final waveform is rendered by the World vocoder. Systems S3 and S4 have same architectures as systems S1 and S2, respectively, but they use the WaveNet vocoder for synthesis. System S5 has the same architecture and vocoder as S4, but instead of using the World vocoder parameters, it predicts the mel-spectrogram as the output.
    \begin{table}[h]
          \caption{Systems developed for experiments.}
         \label{tab:systems}
         \centering
            \begin{tabular}{|c|c|c|c|}
                \hline
                 Sys. ID & TTS model & Ouput & Vocoder  \\ \hline
                 S1 & LSTM     &  MGC+$F0$+BAP+VUV & World \\
                 S2 & Seq2Seq  & " & " \\
                 S3 & LSTM & " & WaveNet \\
                 S4 & Seq2Seq & " & " \\
                 S5 & " & Mel-spectrogram & " \\ \hline
            \end{tabular}
    \end{table}
 
    The baseline S1 system was built as reported in our previous study \cite{bollepalli2017lombard}. We used the fine-tuning method to adapt a LSTM-RNN based TTS system of normal speaking style to Lombard style because this adaptation method showed the best performance. Our previous work used original durations to synthesize Lombard speech. In the current study, however, a separate duration model is built and adapted to Lombard speech. Our Seq2Seq-TTS system is based on the Tacotron-1 architecture \cite{wang2017tacotron} with a few modifications such as predicting the World vocoder parameters instead of the mel-spectrograms as output. 
    Our systems were implemented using an open source repository \footnote{\url{https://github.com/syang1993/gst-tacotron}}. All models were trained on a single GPU with a Nvidia-TITANX 12GB graphics memory card. Before training the model, we processed the data to have even distribution in durations. Because a few utterances are of very long duration and they can slow down training and consume lots of memory, these utterances were removed. The batch size was 32, and 2 acoustic frames were used per each output step. The input linguistic features were mono-phonemes extracted using the Combilex lexicon \cite{richmond2010generating} and represented by one-hot vectors. All acoustic parameters were normalized to have zero mean and unit variance using the standard mean-variance normalization. Linguistic parameters were normalized to lie between 0 and 1 using the min-max normalization.

    \begin{figure}[htbp]
        \centering
        \includegraphics[width=0.9\linewidth, height=2cm]{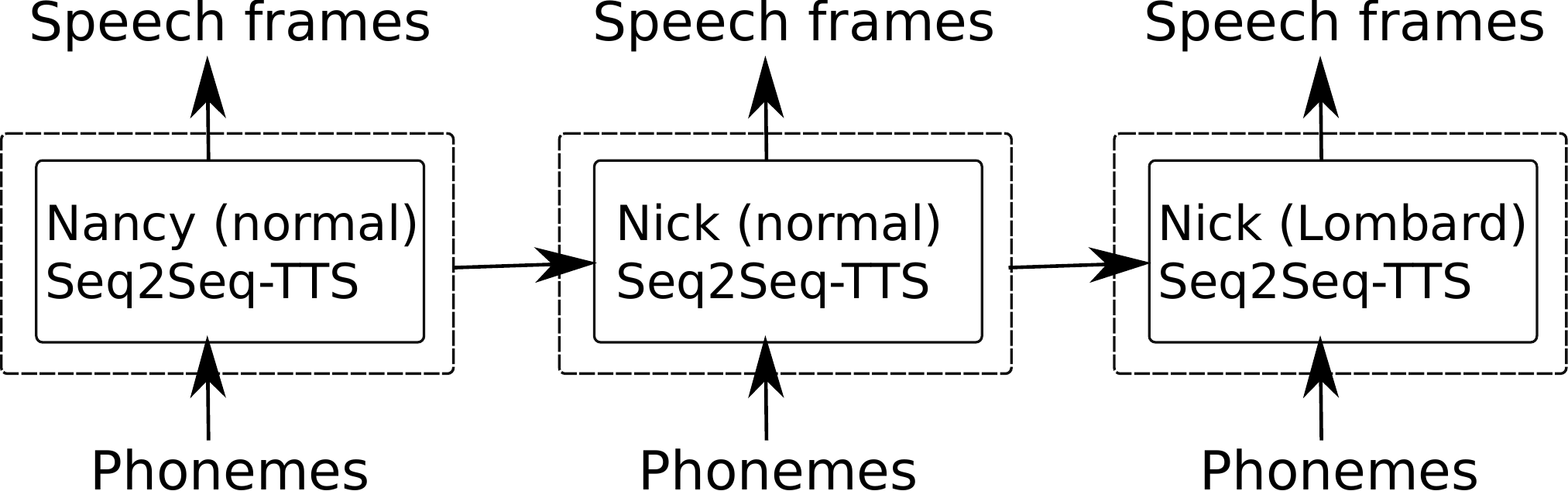}
        \caption{{\it Flow diagram of the adaptation approach.}}
        \label{fig:adapt_block_pic}
    \end{figure}
    
   Figure \ref{fig:adapt_block_pic} illustrates the flow diagram of our adaptation procedure. First, we trained a Seq2Seq-TTS model on the Nancy data of normal style, later that model is adapted to the Nick data of normal style. Then, the Nick normal speech Seq2Seq-TTS model is adapted to Lombard style using the Lombard data of Nick. As seen in Figure~\ref{fig:align_plots}(a), when we trained the Seq2Seq-TTS model using only the Nick data of normal style (i.e., approx. 2 hours of speech), the alignment between the input phoneme sequences and output acoustic frames is not as clear as in Figure~\ref{fig:align_plots}(c), which was obtained by adapting the Nick normal speech data using the Nancy Seq2Seq-TTS model. In informal listening tests, pronunciation errors were perceived when we trained the Seq2Seq-TTS model on the Nick data only; this was most likely because the model was unable to learn adequately with with little data. Thus we decided to train the initial model using the Nancy data (i.e., approx. 16 hours of speech) to learn a good alignment between input phoneme sequences and output acoustic frames.

    \subsubsection{WaveNet vocoder} \label{subsec:wavenet_vocoder}
    We use a WaveNet configuration similar to \cite{Juvela2018}, i.e., three repetitions of a 10-layer convolution stack with exponentially growing dilations, 64 residual channels, and 128 skip channels.
    Separate models were trained for World vocoder acoustic features and mel-spectrograms, using 8-bit categorical cross entropy on quantized $\mu$-law companded signals. 
    %
    We found that excluding BAP from the World features improved performance, so the WaveNet vocoder for World only uses MGCs, VUV and and log$F0$ (interpolated over unvoiced frames). Both the World features and the mel-spectrograms were globally min-max normalized to lie between zero and one.

    \subsection{Subjective evaluation} \label{subsec:subj_eval}
    Two types of listening tests were conducted: 1) speaking style similarity test and 2) comparison category rating (CCR) test of speech naturalness. The goal of the similarity test is to assess whether the technology developed is capable of generating synthetic speech of different speaking styles (normal vs. Lombard) while the CCR test aims to evaluate how much the naturalness of speech is sacrificed when the speaking style is adapted.  We used an evaluation setup similar to \cite{toda2016voice} for the style similarity test. In this evaluation, each stimulus consists of two utterances, the first being a natural speech signal (either normal or Lombard) and the second one a synthesized signal. The subjects were asked to compare the second utterance to the first one and rate the style similarity on a 4-level scale, ranging from 0 {\it(Same: Absolutely sure)} to 4 {\it(Different: Absolutely sure)} \cite{toda2016voice}. In the CCR test, each stimulus consists of a pair of utterances which were stitched together with a silence of 0.5 seconds between them. Subjects were asked to evaluate the naturalness of the second utterance in comparison to the naturalness of the first utterance on a 7-level scale, ranging from -3 {\it(First sample sounds much more natural)} to 3 {\it(Second sample sounds much more natural)}.

    Both tests were conducted on FigureEight \footnote{\url{https://www.figure-eight.com/}}, a crowdsource platform (see \cite{Juvela2018} for more details of conducting the tests). We selected 16 utterances randomly from the test set for each system. Each utterance was evaluated by 50 listeners, and the listeners were screened using natural reference null pairs and artificially corrupted anchor samples.

    \subsection{Results} \label{subsec:results}
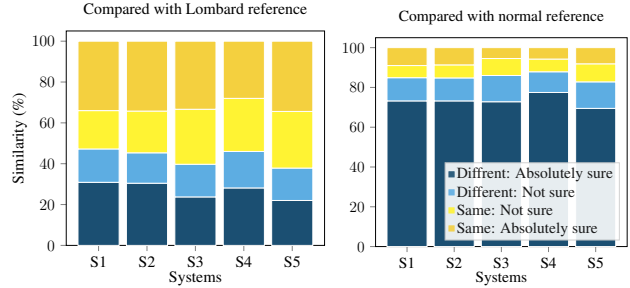
\begin{figure}[htpb]
    \large
    {
    \resizebox*{.24\textwidth}{!}{
\begin{tikzpicture}

\definecolor{color0}{rgb}{0.105882352941176,0.309803921568627,0.447058823529412}
\definecolor{color1}{rgb}{0.364705882352941,0.67843137254902,0.886274509803922}
\definecolor{color2}{rgb}{1,0.952941176470588,0.2}
\definecolor{color3}{rgb}{0.956862745098039,0.815686274509804,0.247058823529412}

\begin{axis}[
legend cell align={left},
legend style={at={(0.97,0.03)}, anchor=south east, draw=white!80.0!black},
tick align=outside,
tick pos=left,
title={Compared with Lombard reference},
x grid style={white!69.01960784313725!black},
xlabel={Systems},
xmin=-0.6675, xmax=4.6675,
xtick={0,1,2,3,4},
xticklabels={S1,S2,S3,S4,S5},
y grid style={white!69.01960784313725!black},
ylabel={Similarity (\%)},
ymin=0, ymax=105
]
\draw[draw=white,fill=color0] (axis cs:-0.425,0) rectangle (axis cs:0.425,30.8888888888889);
\draw[draw=white,fill=color0] (axis cs:0.575,0) rectangle (axis cs:1.425,30.4444444444444);
\draw[draw=white,fill=color0] (axis cs:1.575,0) rectangle (axis cs:2.425,23.7333333333333);
\draw[draw=white,fill=color0] (axis cs:2.575,0) rectangle (axis cs:3.425,28.1333333333333);
\draw[draw=white,fill=color0] (axis cs:3.575,0) rectangle (axis cs:4.425,22);
\draw[draw=white,fill=color1] (axis cs:-0.425,30.8888888888889) rectangle (axis cs:0.425,47.2222222222222);
\draw[draw=white,fill=color1] (axis cs:0.575,30.4444444444444) rectangle (axis cs:1.425,45.3333333333333);
\draw[draw=white,fill=color1] (axis cs:1.575,23.7333333333333) rectangle (axis cs:2.425,39.7333333333333);
\draw[draw=white,fill=color1] (axis cs:2.575,28.1333333333333) rectangle (axis cs:3.425,46);
\draw[draw=white,fill=color1] (axis cs:3.575,22) rectangle (axis cs:4.425,37.8666666666667);
\draw[draw=white,fill=color2] (axis cs:-0.425,47.2222222222222) rectangle (axis cs:0.425,66);
\draw[draw=white,fill=color2] (axis cs:0.575,45.3333333333333) rectangle (axis cs:1.425,65.7777777777778);
\draw[draw=white,fill=color2] (axis cs:1.575,39.7333333333333) rectangle (axis cs:2.425,66.6666666666667);
\draw[draw=white,fill=color2] (axis cs:2.575,46) rectangle (axis cs:3.425,72);
\draw[draw=white,fill=color2] (axis cs:3.575,37.8666666666667) rectangle (axis cs:4.425,65.6);
\draw[draw=white,fill=color3] (axis cs:-0.425,66) rectangle (axis cs:0.425,100);
\draw[draw=white,fill=color3] (axis cs:0.575,65.7777777777778) rectangle (axis cs:1.425,100);
\draw[draw=white,fill=color3] (axis cs:1.575,66.6666666666667) rectangle (axis cs:2.425,100);
\draw[draw=white,fill=color3] (axis cs:2.575,72) rectangle (axis cs:3.425,100);
\draw[draw=white,fill=color3] (axis cs:3.575,65.6) rectangle (axis cs:4.425,100);
\end{axis}

\end{tikzpicture}} 
    \resizebox*{.22\textwidth}{!}{
\begin{tikzpicture}

\definecolor{color0}{rgb}{0.105882352941176,0.309803921568627,0.447058823529412}
\definecolor{color1}{rgb}{0.364705882352941,0.67843137254902,0.886274509803922}
\definecolor{color2}{rgb}{1,0.952941176470588,0.2}
\definecolor{color3}{rgb}{0.956862745098039,0.815686274509804,0.247058823529412}

\begin{axis}[
legend cell align={left},
legend entries={{Diffrent: Absolutely sure},{Different: Not sure},{Same: Not sure},{Same: Absolutely sure}},
legend style={at={(0.97,0.03)}, anchor=south east, draw=white!80.0!black, opacity=0.9},
tick align=outside,
tick pos=left,
title={Compared with normal reference},
x grid style={white!69.01960784313725!black},
xlabel={Systems},
xmin=-0.6675, xmax=4.6675,
xtick={0,1,2,3,4},
xticklabels={S1,S2,S3,S4,S5},
y grid style={white!69.01960784313725!black},
ymin=0, ymax=105
]
\draw[draw=white,fill=color0] (axis cs:-0.425,0) rectangle (axis cs:0.425,73.2222222222222);
\draw[draw=white,fill=color0] (axis cs:0.575,0) rectangle (axis cs:1.425,73.2222222222222);
\draw[draw=white,fill=color0] (axis cs:1.575,0) rectangle (axis cs:2.425,72.8);
\draw[draw=white,fill=color0] (axis cs:2.575,0) rectangle (axis cs:3.425,77.4666666666667);
\draw[draw=white,fill=color0] (axis cs:3.575,0) rectangle (axis cs:4.425,69.4666666666667);
\addlegendimage{only marks, mark=square*, color=color0};
\draw[draw=white,fill=color1] (axis cs:-0.425,73.2222222222222) rectangle (axis cs:0.425,84.8888888888889);
\draw[draw=white,fill=color1] (axis cs:0.575,73.2222222222222) rectangle (axis cs:1.425,84.7777777777778);
\draw[draw=white,fill=color1] (axis cs:1.575,72.8) rectangle (axis cs:2.425,86);
\draw[draw=white,fill=color1] (axis cs:2.575,77.4666666666667) rectangle (axis cs:3.425,87.8666666666667);
\draw[draw=white,fill=color1] (axis cs:3.575,69.4666666666667) rectangle (axis cs:4.425,82.8);
\addlegendimage{only marks, mark=square*, color=color1};
\draw[draw=white,fill=color2] (axis cs:-0.425,84.8888888888889) rectangle (axis cs:0.425,91);
\draw[draw=white,fill=color2] (axis cs:0.575,84.7777777777778) rectangle (axis cs:1.425,91.3333333333333);
\draw[draw=white,fill=color2] (axis cs:1.575,86) rectangle (axis cs:2.425,94.5333333333333);
\draw[draw=white,fill=color2] (axis cs:2.575,87.8666666666667) rectangle (axis cs:3.425,94.2666666666667);
\draw[draw=white,fill=color2] (axis cs:3.575,82.8) rectangle (axis cs:4.425,91.8666666666667);
\addlegendimage{only marks, mark=square*, color=color2};
\draw[draw=white,fill=color3] (axis cs:-0.425,91) rectangle (axis cs:0.425,100);
\draw[draw=white,fill=color3] (axis cs:0.575,91.3333333333333) rectangle (axis cs:1.425,100);
\draw[draw=white,fill=color3] (axis cs:1.575,94.5333333333333) rectangle (axis cs:2.425,100);
\draw[draw=white,fill=color3] (axis cs:2.575,94.2666666666667) rectangle (axis cs:3.425,100);
\draw[draw=white,fill=color3] (axis cs:3.575,91.8666666666667) rectangle (axis cs:4.425,100);
\addlegendimage{only marks, mark=square*, color=color3};
\end{axis}

\end{tikzpicture}}
    }
    \caption{Results of the style similarity test.}
    \label{fig:styletest_vocoder}
    \end{figure}

    The results of the style similarity test are plotted in Figure~\ref{fig:styletest_vocoder}. From the right pane of the figure, it can be observed that synthesized speech by all adapted systems were rated to sound different from natural normal speech with high confidence. When compared to the natural Lombard reference (left pane), system S5 was rated highest, followed by systems S3, S4, S2, and S1. System S5 was built using the Seq2Seq-TTS model and the mel-spectrogram as output. It can be clearly seen that those systems that employed the WaveNet vocoder got higher scores than the ones that used the more traditional World vocoder. Further, system S5, which is based on conditioning the WaveNet vocoder with mel-spectrograms, got a higher score than the ones that used the World vocoder, confirming the findings made in a earlier study \cite{adiga2018use}.  From this results we can conclude that even though we trained the WaveNet vocoder with only 30 minutes of Lombard speech, system S5 generated synthetic speech that was most Lombard-like among the systems compared. 

\begin{figure}[htpb]
        \centering




\begin{tikzpicture}[x=1pt,y=1pt]
\definecolor{fillColor}{RGB}{255,255,255}
\path[use as bounding box,fill=fillColor,fill opacity=0.00] (0,0) rectangle (108.41,144.54);
\begin{scope}
\path[clip] (  0.00,  0.00) rectangle (108.41,144.54);
\definecolor{drawColor}{RGB}{0,0,0}

\path[draw=drawColor,line width= 0.4pt,line join=round,line cap=round] ( 54.19,111.68) circle (  1.80);

\path[draw=drawColor,line width= 0.4pt,line join=round,line cap=round] ( 54.19, 18.84) circle (  1.80);

\path[draw=drawColor,line width= 0.4pt,line join=round,line cap=round] ( 54.19, 36.90) circle (  1.80);

\path[draw=drawColor,line width= 0.4pt,line join=round,line cap=round] ( 54.19, 44.41) circle (  1.80);

\path[draw=drawColor,line width= 0.4pt,line join=round,line cap=round] ( 48.00, 14.24) -- ( 48.00,130.30);

\path[draw=drawColor,line width= 0.4pt,line join=round,line cap=round] ( 48.00, 14.24) -- ( 43.20, 14.24);

\path[draw=drawColor,line width= 0.4pt,line join=round,line cap=round] ( 48.00, 33.58) -- ( 43.20, 33.58);

\path[draw=drawColor,line width= 0.4pt,line join=round,line cap=round] ( 48.00, 52.93) -- ( 43.20, 52.93);

\path[draw=drawColor,line width= 0.4pt,line join=round,line cap=round] ( 48.00, 72.27) -- ( 43.20, 72.27);

\path[draw=drawColor,line width= 0.4pt,line join=round,line cap=round] ( 48.00, 91.61) -- ( 43.20, 91.61);

\path[draw=drawColor,line width= 0.4pt,line join=round,line cap=round] ( 48.00,110.96) -- ( 43.20,110.96);

\path[draw=drawColor,line width= 0.4pt,line join=round,line cap=round] ( 48.00,130.30) -- ( 43.20,130.30);

\node[text=drawColor,anchor=base east,inner sep=0pt, outer sep=0pt, scale=  0.80] at ( 38.40, 11.49) {-1.0};

\node[text=drawColor,anchor=base east,inner sep=0pt, outer sep=0pt, scale=  0.80] at ( 38.40, 30.83) {-0.5};

\node[text=drawColor,anchor=base east,inner sep=0pt, outer sep=0pt, scale=  0.80] at ( 38.40, 50.17) {0.0};

\node[text=drawColor,anchor=base east,inner sep=0pt, outer sep=0pt, scale=  0.80] at ( 38.40, 69.52) {0.5};

\node[text=drawColor,anchor=base east,inner sep=0pt, outer sep=0pt, scale=  0.80] at ( 38.40, 88.86) {1.0};

\node[text=drawColor,anchor=base east,inner sep=0pt, outer sep=0pt, scale=  0.80] at ( 38.40,108.20) {1.5};

\node[text=drawColor,anchor=base east,inner sep=0pt, outer sep=0pt, scale=  0.80] at ( 38.40,127.54) {2.0};

\node[text=drawColor,rotate= 90.00,anchor=base,inner sep=0pt, outer sep=0pt, scale=  0.80] at ( 17.28, 72.27) {Score difference};

\path[draw=drawColor,line width= 0.4pt,line join=round,line cap=round] ( 54.19,108.91) -- ( 54.19,114.45);

\path[draw=drawColor,line width= 0.4pt,line join=round,line cap=round] ( 50.57,108.91) --
	( 54.19,108.91) --
	( 57.80,108.91);

\path[draw=drawColor,line width= 0.4pt,line join=round,line cap=round] ( 57.80,114.45) --
	( 54.19,114.45) --
	( 50.57,114.45);

\path[draw=drawColor,line width= 0.4pt,line join=round,line cap=round] ( 54.19, 15.88) -- ( 54.19, 21.80);

\path[draw=drawColor,line width= 0.4pt,line join=round,line cap=round] ( 50.57, 15.88) --
	( 54.19, 15.88) --
	( 57.80, 15.88);

\path[draw=drawColor,line width= 0.4pt,line join=round,line cap=round] ( 57.80, 21.80) --
	( 54.19, 21.80) --
	( 50.57, 21.80);

\path[draw=drawColor,line width= 0.4pt,line join=round,line cap=round] ( 54.19, 33.85) -- ( 54.19, 39.96);

\path[draw=drawColor,line width= 0.4pt,line join=round,line cap=round] ( 50.57, 33.85) --
	( 54.19, 33.85) --
	( 57.80, 33.85);

\path[draw=drawColor,line width= 0.4pt,line join=round,line cap=round] ( 57.80, 39.96) --
	( 54.19, 39.96) --
	( 50.57, 39.96);

\path[draw=drawColor,line width= 0.4pt,line join=round,line cap=round] ( 54.19, 41.02) -- ( 54.19, 47.80);

\path[draw=drawColor,line width= 0.4pt,line join=round,line cap=round] ( 50.57, 41.02) --
	( 54.19, 41.02) --
	( 57.80, 41.02);

\path[draw=drawColor,line width= 0.4pt,line join=round,line cap=round] ( 57.80, 47.80) --
	( 54.19, 47.80) --
	( 50.57, 47.80);

\node[text=drawColor,anchor=base west,inner sep=0pt, outer sep=0pt, scale=  0.80] at ( 62.33,109.84) {Ref};

\node[text=drawColor,anchor=base west,inner sep=0pt, outer sep=0pt, scale=  0.80] at ( 62.33, 17.00) {S1};

\node[text=drawColor,anchor=base west,inner sep=0pt, outer sep=0pt, scale=  0.80] at ( 62.33, 35.07) {S3};

\node[text=drawColor,anchor=base west,inner sep=0pt, outer sep=0pt, scale=  0.80] at ( 62.33, 42.57) {S5};

\end{scope}
\end{tikzpicture}

        \caption{Combined score differences obtained from the CCR test of naturalness. Error bars are t-statistic based 95\% confidence intervals for the mean.}
        \label{fig:ccr_results}
    \end{figure}
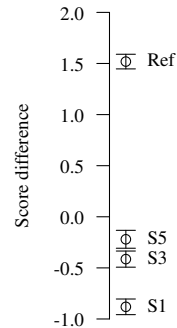

    For the CCR test, we included only systems S1, S3 and S5. System S1 can be regarded as the baseline. System S3 was selected because it was the best system in the similarity test with the LSTM models and the WaveNet vocoder. S5 was selected because it was the best system in overall in the similarity test. 
    The results of the CCR test are shown in Figure~\ref{fig:ccr_results}. The scores were calculated by reordering the ratings for each system and pooling together all ratings the system received. Natural Lombard speech was included in the tests as a reference system. The plot show mean ratings with 95\% confidence, corrected for multiple comparisons. As expected, the Lombard reference signal was rated highest followed by S5, S3 and S1. System S5 got a significant better score than the baseline system S1 and the LSTM based system S3. 

  \section{Conclusions} \label{sec:conclusions}
  This paper compared different TTS models and vocoders to adapt the speaking style of speech synthesis from normal to Lombard. The study proposes using an adaptation method based on fine-tuning combined with sequence-to-sequence based TTS models and the WaveNet vocoder conditioned using mel-spectrograms. Listening tests show that the proposed method outperformed the previous best method that was developed using a LSTM-RNN based adapted system.
  Future work includes an extensive subjective evaluations and training both the WaveNet and Seq2Seq-TTS model in a single pipeline. 

 \section{Acknowledgements} \label{sec:acknowledments}
 The study was funded by the Academy of Finland (project 312490).
  We  thank Vassilis Tsiaras for sharing his WaveNet implementation. 


{
\bibliographystyle{IEEEtran}
\bibliography{refs}
}

\end{document}